\documentclass[12pt]{article}
\pdfoutput = 1

\usepackage{amsfonts}
\usepackage{amsmath}
\usepackage{amssymb}
\usepackage{appendix}
\usepackage{braket}
\usepackage{enumitem}
\usepackage{float}
\usepackage{graphicx}
\usepackage{mathabx}
\usepackage[mathcal]{eucal}
\usepackage{mathrsfs}
\usepackage{pifont}
\usepackage{setspace}
\usepackage{slashed}
\usepackage{subcaption}
\usepackage{titlesec}
\usepackage{tikz}
\usetikzlibrary{arrows}
\usetikzlibrary{arrows.meta}
\usetikzlibrary{patterns}

\usepackage[margin = 2.5cm]{geometry}
\usepackage{color}
\definecolor{darkblue}{rgb}{0,0,0.6}
\definecolor{purple}{rgb}{0.4,.2,0.7}
\definecolor{darkgreen}{rgb}{0,0.5,0}

\usepackage[bookmarks = false, colorlinks = true, linkcolor = darkblue, citecolor = purple]{hyperref}

\renewcommand{\i}{\mathrm{i}}
\renewcommand{\d}{\mathrm{d}}
\DeclareMathOperator{\Tr}{Tr}

\newcommand{\sgen}{S_\text{gen}}

\begin{document}

\thispagestyle{empty}
\begin{center}
    ~\vspace{5mm}
    
    {\Large 

      Lectures on Quantum Extremal Surfaces and the Page Curve
    
    }
    
    \vspace{0.4in}
    
    {\large Raghu Mahajan$^{1,2}$}

    \vspace{0.4in}

	$^1$ International Center for Theoretical Sciences,  Shivakote Village,  Hesaraghatta Hobli,  Bengaluru 560089,  Karnataka,  India
    \vskip1ex
    $^2$ Department of Physics, Stanford University, Stanford, CA 94305-4060, USA 

    \vspace{0.1in}
    
    {\tt raghu.mahajan@icts.res.in}
\end{center}

\vspace{0.4in}

\begin{abstract}

This article is a write-up of pedagogical lectures delivered at 
the Asia Pacific Center for Theoretical Physics online winter school held in January 2021, and also as part of the Quantum Information, Quantum Field Theory and Gravity program held at ICTS,  Bengaluru,  in August 2024.
The topics covered include a brief derivation of Hawking radiation from the perspective of correlation functions,  a description of entropy paradoxes in the eternal black hole,  and details of the associated entanglement island computations in two-dimensional models.

\end{abstract}

\pagebreak

\tableofcontents


\section{Introduction}
Einstein's theory of general relativity describes classical gravity \cite{Misner:1973prb, Wald:1984rg}.
The equations of motion of general relativity,  also called Einstein's equations,  determine the geometry of spacetime from appropriate initial conditions.
Needless to say,  the theory is extremely well-tested experimentally \cite{Will:2014kxa}.
It is known that gravitational collapse of stars can result in the formation of black holes \cite{Oppenheimer:1939ue}.
Black holes are solutions to Einstein's equations which have horizons.
Horizons are co-dimension one null surfaces in spacetime which have the property that light emitted from the region behind the surface cannot make it to infinity.
Without any stress-energy sources,  a black hole is completely characterized by its mass and angular momentum.
A rotating black holes is called a Kerr black hole,  and it's non-rotating limit is known as the Schwarzschild black hole.
The detection of gravitational waves from the merger of a pair of black holes \cite{LIGOScientific:2016aoc} and the results from the Event Horizon Telescope \cite{EventHorizonTelescope:2019dse} have established beyond doubt the existence of black holes in our universe.

It is natural to expect that classical general relativity will arise as the limit of a quantum theory of gravity.
A quantum theory of gravity applicable to our universe continues to evade us,  but string theory is currently the most serious candidate \cite{Polchinski:1998rq, Polchinski:1998rr}.
The full power of quantum gravity would be needed to address physics questions that are sensitive to singularities in general relativity.\footnote{Concretely,  we can think about either the singularity inside black holes or the big bang singularity.}
Given that the exact nature of a full theory of quantum gravity remains unknown,  we can ask whether there are interesting phenomena that happen when the quantum effects are, in some sense,  ``milder" than the effects near singularities.

It is a celebrated result of Hawking \cite{Hawking:1974rv,  Hawking:1975vcx} that,  when quantum effects are taken into account,  black holes emit particles.
This emitted gas of particles has a nonzero temperature,  and has come to be referred to as Hawking radiation.
For a non-rotating black hole with horizon radius $r_+$,  this temperature $T_H$ is given by the following expression:
\begin{align}
 k_B T_H = \hbar \cdot \frac{c}{r_+} \cdot \frac{1}{4\pi}\, .
 \label{thawk}
\end{align}
Note that the temperature goes to zero in the classical limit $\hbar \to 0$.
Another important fact is that this effect is intimately tied to the existence of horizons.

There is another remarkable fact about black holes.
This is the fact that black holes have an entropy,  in the sense of statistical physics,  which is equal to the area of the horizon in Planck units \cite{Bekenstein:1973ur,  Bardeen:1973gs,Gibbons:1976ue}
\begin{align}
S_{\text{BH}} = k_B \cdot \frac{1}{4} \cdot 4\pi r_+^2 \cdot \frac{c^3}{G\hbar}\, .
\label{sbek}
\end{align}
In other words,  the black hole spacetime should be thought of as an ensemble of a large number of microscopic states,  with the number of states being $\exp(S_{\text{BH}})$.
The fact that this entropy is finite and devoid of additive ambiguities is another imprint of the underlying quantum theory of gravity.\footnote{It is not directly relevant to our considerations here,  but it is one of the remarkable successes of string theory that it can precisely reproduce the area-entropy relation of certain special but large class of black holes from a counting of microscopic configurations of strings and membranes \cite{Strominger:1996sh,Sen:2007qy}.}

Let us take the initial star that undergoes gravitational collapse to form the black hole to be in a zero-entropy state.
The emission of Hawking radiation by the black hole causes it to lose mass,  due to which the black hole horizon starts to shrink in size,  and eventually disappears completely---it ``evaporates."
At this point,  something very perplexing has happened: The black hole has disappeared,  and so the initial star that was in a zero-entropy state has been completely converted into a gas of particles with a nonzero amount of entropy (since particles with nonzero temperature have nonzero entropy). 
This would lead us to the disturbing conclusion that the time evolution process that takes the initial star into the final Hawking radiation (with no black hole) is not reversible,  at complete odds with the rest of our knowledge about the physical world \cite{Hawking:1976ra}.
This seeming paradox has come to be known as the ``black hole information paradox," or 
the ``black hole information problem."

This result has ignited a lot of heated debate in the roughly fifty years that we have known about it.
This is not the place to review the full history of how our understanding of this paradox evolved. 
We will take a pedagogical approach and simply state the modern perspective.
\begin{enumerate}
\item Time evolution processes,  whether or not they involve the evaporation of black holes,  are unitary,  and hence,  reversible.
This is supported by the AdS/CFT correspondence \cite{Maldacena:1997re},  which tells us that black holes in the presence of a negative cosmological constant are completely equivalent to a plasma of quarks and gluons in a nonabelian gauge theory without gravity \cite{Witten:1998zw,  Maldacena:2001kr}.\footnote{The nonabelian gauge theory is defined on the boundary of the black hole spacetime.}
Certainly,  the time evolution in a nonabelian gauge theory without gravity is reversible.

\item It is not useful to think about the black hole information paradox at only the final stages of black hole evaporation.
The horizon radius is then very small and we would have to think about the full theory of quantum gravity.
Luckily,  a paradox exists even when the black hole has shrunk by only an order one fraction (larger than one-half).
This is the result of Page \cite{Page:1993df,  Page:1993wv,Page:2013dx}.
The Page version of the paradox stems from the fact that,  for a quantum many body system in a generic pure state,  the bipartite von Neumann entropy of a subsystem and its complement are equal,  and equal to the number of degrees of freedom in the \emph{smaller} subsystem.
Let us take the two subsystems to be (a) the black hole,  at a time which is some order one fraction of its lifetime,  and (b) the Hawking radiation that has been emitted up to that point of time.
On one hand,  Hawking's result says that the entropy of the emitted radiation should be a monotonically increasing,  almost linear function of time. 
On the other hand,  the above result of Page suggests that after the half-way evaporation stage,  the entropy of Hawking radiation should start going down and be equal to the area-entropy of the black hole,  which is decreasing with time.
This leads to a contradiction not just in the final-stages of the black hole evaporation,  but also when the black hole is still large. 
This makes the problem amenable to a semi-classical treatment,  which is a huge step forward for a deeper understanding of the problem.
Given the importance of this result,  the time when the black-hole has evaporated half-way is referred to as the ``Page time," and the tent-shaped time-dependence of the entropy of Hawking radiation is referred to as the ``Page curve."

\item Any question about quantum theory of gravity must be formulated as a path integral problem with boundary conditions that are appropriate to the question being asked.
If we want to compute the von Neumann entropy of Hawking radiation from first principles,  we must first find a way to formulate the question as a path integral with the correct boundary conditions.
This is possible via the replica technique \cite{Lewkowycz:2013nqa},  which is a generalization of the fundamental work of Gibbons and Hawking \cite{Gibbons:1976ue}.
Since this point has been the source of innumerable confusions,  we cannot emphasize enough that the replica method is deployed on the boundary,  and not in bulk of the gravitational spacetime.
Since we are asking questions in a regime where semi-classical methods are valid,  we can now compute the path integral by the saddle point method.
Two fundamental questions are: What are the relevant saddle points and what are the values of their on-shell action?
\item When one formulates the question this way using first principles,  one discovers that,  at early times, the standard black hole saddle point dominates.
This leads to the growth of the entropy of Hawking radiation according to Hawking's prediction.
However,  after the Page time,  a different saddle point starts to have lower action and thus dominates the answer.
This saddle point is a ``replica wormhole" \cite{Penington:2019kki,  Almheiri:2019qdq},  which is a bulk geometry that connects the various replicas of the boundary that enter the replica technique. 
The on-shell action of this new saddle point leads to an entropy that tracks the shrinking area of the black hole,  hence reproducing the downward slope of the Page curve after the Page time.
The existence of this new saddle-point is highly nontrivial and follows a series of developments in the computation of von Neumann entropies in the context of the AdS/CFT: the Ryu-Takayanagi (RT) formula \cite{Ryu:2006bv},  the one-loop correction to the RT formula \cite{Faulkner:2013ana},  the quantum extremal surface prescription \cite{Engelhardt:2014gca},  and a path integral understanding of these results \cite{Lewkowycz:2013nqa}.
The Page curve was derived using the quantum extremal surface prescription in the remarkable papers \cite{Penington:2019npb,  Almheiri:2019psf}. 
These two papers in turn motivated the works \cite{Penington:2019kki,  Almheiri:2019qdq} which identified the replica-wormhole saddle points in the path integral setup of entropy computations.
\end{enumerate}

The structure of this article is as follows. 
We will first present a derivation of the phenomenon of Hawking radiation using a lesser known technique \cite{Fredenhagen:1989kr} that focuses on the KMS property of thermal two-point correlators as opposed to the usual method using Bogoliubov coefficients \cite{Hawking:1975vcx}.
Then we will discuss some necessary formulas for von Neumann entropy in two-dimensional conformal field theories.
We will then introduce our model of an eternal two-dimensional black hole in JT gravity \cite{Maldacena:2016upp} coupled to a two-dimensional conformal field theory.
Following \cite{Mathur:2014dia,  Almheiri:2019yqk},  we will discuss the version of the entropy paradox and the Page curve in this eternal black-hole setup (as opposed to a black hole formed from collapse).
We will then use the quantum extremal surface prescription \cite{Engelhardt:2014gca} to derive the Page curve via a competition between two candidate extrema.
We restrict to the case of an eternal black-hole since it is simpler compared to an evaporating black hole but contains all the necessary ingredients.
We will not discuss replica-wormholes in detail,  see the review \cite{Almheiri:2020cfm}, but we hope that we have conveyed the basic idea in the introduction.

\section{The KMS property and the thermality of Hawking radiation}
\subsection{The KMS property}

Consider a quantum system in the thermal equilibrium density matrix at inverse-temperature $\beta>0$.
We define a strip region in the complex time plane
\begin{align}
S := \{
t \in \mathbb{C} : -\beta < \text{Im}(t) < 0
\}\, .
\end{align}
In other words,  we consider a horizontal open strip lying between the lines $y=0$ and $y=-\beta$.
Let $\phi$ be an operator in the quantum system under consideration.
For $t \in S$,  we define the correlation function
\begin{align}
G(t) &:= \frac{1}{Z} \Tr \left( e^{-\beta H} \phi(t) \phi(0) \right)\quad \quad \quad 
(\text{for } t  \in S) \\
&= \frac{1}{Z} \Tr \left( e^{-\beta H} e^{+\i H t} \phi\, e^{-\i H t} \phi \right)\, , \label{gtdef}\\
\text{where } Z &:= \Tr e^{-\beta H}\, .
\end{align}
Here $H$ is the Hamiltonian,  or the operator that generates translations of the time coordinate $t$.
Note that the first exponential in (\ref{gtdef}) is well defined since $-\beta + \i t = - \beta - \text{Im}(t) + \i \text{Re}(t)$ has negative imaginary part (recall that $t\in S$ and hence $\text{Im}(t) > - \beta$).
Similarly,  the second exponential is well defined since $-\i t = \text{Im}(t) - \i \text{Re}(t)$ also has negative imaginary part.
In other words,  $G(t)$ is an analytic function in the strip $S$.
In most systems of interest,  $G(t)$ has interesting singularities on the boundary of $S$.

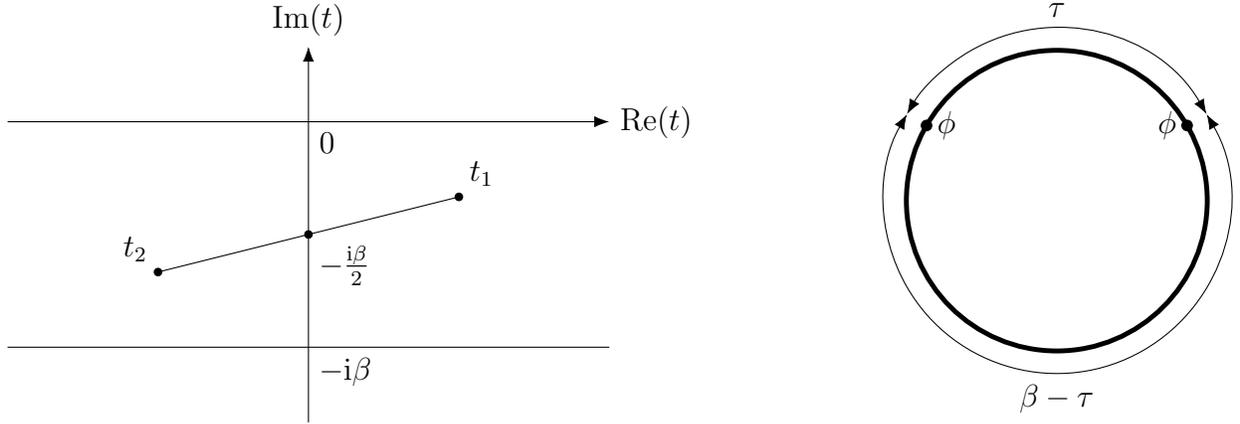
\begin{figure}
\centering
\begin{tikzpicture}
\draw[-{Latex[length=2mm]}] (0,-4) -- (0,1) node[above] {$\text{Im}(t)$};
\draw[-{Latex[length=2mm]}] (-4,0) -- (4,0) node[right] {$\text{Re}(t)$};
\draw[-] (-4,-3) -- (4,-3);

\draw[fill] (2,-1) circle [radius=0.05] node[above right] {$t_1$};
\draw[fill] (-2,-2) circle [radius=0.05] node[above left] {$t_2$};
\draw[-] (2,-1) -- (-2,-2);
\draw[fill] (0,-1.5) circle [radius=0.05];
\node[below right] at (0,-1.5) {$-\frac{\i\beta}{2}$};

\node[below right] at (0,-3) {$-\i\beta$};
\node[below right] at (0,0) {$0$};
\end{tikzpicture}
  \hfill
\begin{tikzpicture}
\draw[line width=0.65mm] (0,0) circle (2);

\filldraw (30:2) circle (2pt) node[left] {$\phi$};
\filldraw (150:2) circle (2pt) node[right] {$\phi$};

\draw[{Latex[length=2mm]}-{Latex[length=2mm]}] (30:2.3) arc[start angle=30, end angle=150, radius=2.3] node[midway, above] {$\tau$};

\draw[{Latex[length=2mm]}-{Latex[length=2mm]}] (30:2.3) arc[start angle=30, end angle=-210, radius=2.3] node[midway, below] {$\beta - \tau$};
\end{tikzpicture}

\caption{The KMS property (for identical scalar Grassmann-even operators).  The thermal correlator $\langle \phi(t)\phi(0) \rangle$ is analytic in the strip $-\beta < \text{Im}(t) < 0$.  The KMS property states that $G(t_1) = G(t_2)$ if $\frac{t_1 + t_2}{2} = \frac{-\i\beta}{2} $.
If $t$ is purely imaginary,  say $t = - \i\tau$,  then the correlator can be represented via a purely Euclidean path integral.
The KMS property is then the statement that the separation between the two operators can equally well be taken to be $\tau$  or $\beta - \tau$.
}
\label{figkms}
\end{figure}

The Kubo-Martin-Schwinger (KMS) property \cite{kubo1957statistical, martin1959theory} states that
\begin{align}
G(t_1) = G(t_2) \quad \text{ if }\quad  \frac{t_1 + t_2}{2}= - \i \, \frac{\beta}{2}\, .
\label{kms}
\end{align}
This follows from the definition of $G(t_1)$ in (\ref{gtdef}) since using $\beta = \i (t_1 + t_2)$,  we can rewrite the numerator of (\ref{gtdef}) as $\Tr (e^{-\i H t_2} \phi \,e^{-\i H t_1 }\phi)$.
This latter expression is symmetric under the exchange of $t_1$ and $t_2$ because of the cyclic property of the trace.
Geometrically,  the KMS property (\ref{kms}) says that the function $G(t)$ is equal at two points in $S$ such that the mid-point of the line segment joining those two points coincides with the ``mid-point" $-\i \frac{\beta}{2}$ of the strip; see the left panel of figure \ref{figkms}.
If we evaluate the correlator at a purely imaginary value of $t$,  say $t = - \i \tau$ with $0 < \tau < \beta$,  the KMS property (\ref{kms}) says that the correlator is invariant under $\tau \leftrightarrow \beta - \tau$; see the right panel of figure \ref{figkms}.

There are straightforward generalizations of the KMS property when the two operators are not identical,  or when the operators are Grassmann odd.

The KMS property is an alternative \emph{equivalent} characterization of thermal states \cite{Araki:1977px,sewell1977kms}.
This is useful conceptually since quantum systems usually encountered in quantum field theory do not have a well-defined density matrix.
So,  it is not rigorous to define the thermal state of a general quantum system by declaring its density matrix to be $e^{-\beta H}$.
It is thus conceptually cleaner to define a thermal state via the fact that correlation functions in a thermal state satisfy the KMS property.

\subsection{KMS property of Rindler correlators in the Minkowski vacuum}
\begin{figure}
\centering
\begin{tikzpicture}

\draw[-{Latex[length=2mm]}] (0,-4) -- (0,4) node[above] {$x^0$};
\draw[-{Latex[length=2mm]}] (-4,0) -- (4,0) node[right] {$x^1$};

\draw[domain=-1.6:1.6, samples=100, smooth, variable=\t] 
        plot ({1.5*cosh(\t)}, {1.5*sinh(\t)});

\draw[dashed] (0,0) -- (4,4);
\draw[dashed] (0,0) -- (4,-4);

\filldraw (1.5,0) circle (2pt) node[below right] {$\phi$};
\filldraw (2.31462,1.7628) circle (2pt) node[right] {$\phi$};

\end{tikzpicture}
\caption{The right Rindler wedge is the region $R^+$ obeying the inequality $x^1 > \vert x^0 \vert$.  A curve of constant $\xi$ is depicted,  it is a timelike trajectory with constant proper acceleration.  We consider the two-point function of an operator in the Minkowski vacuum state,  with the two-points separated purely in the $\eta$ direction.  This correlator obeys the KMS property with $\beta = 2\pi$ and with respect to the generator of $\eta$-translations.}
\label{figrindler}
\end{figure}
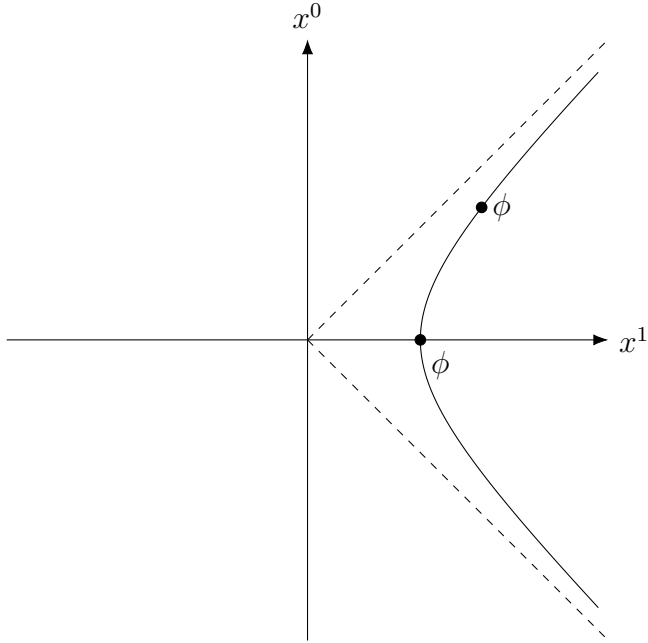

In this subsection we want to discuss the physics of the Rindler wedge,  vis-a-vis the physics of Minkowski space.
This is  a simple example which is very useful in the understanding of Hawking radiation.
Our presentation will focus on the important ideas that are needed later.
We refer the reader to the extremely detailed and pedagogical review article \cite{Takagi:1986kn} for a more information.
Some original references are \cite{Rindler:1966zz,  Unruh:1976db,Bisognano:1976za}.

We denote $D$-dimensional Minkowski space by $\mathbb{R}^{1,D-1}$.
We use coordinates $(x^0,\ldots, x^{D-1})$ with the line element
\begin{align}
\d s^2 = - (\d x^0)^2 + (\d x^1)^2 + \ldots + (\d x^{D-1})^2\, .
\end{align}
The right Rindler wedge is defined as the subset
\begin{align}
R^+ := \{
(x^0,\ldots, x^{D-1}) \in \mathbb{R}^{1,D-1} : x^1 > \vert x^0 \vert
\}\, .
\label{minkds2}
\end{align}
Note that it is defined as an open set.
Let us explain the significance of $R^+$.

Consider a quantum field theory on $\mathbb{R}^{1,D-1}$.
Full Minkowski space has a Cauchy hypersurface defined simply by $x^0 = 0$.
If we restrict ourselves to half of this hypersurface defined by $x^1 > 0$,  then the region $R^+$ is the domain of dependence of this half-Cauchy surface.
If we restrict states of quantum fields in full Minkowski space to $R^+$,  we will typically obtain mixed states since we are effectively tracing out degrees of freedom in the other half of the Cauchy hypersurface.
A basic,  yet deep fact is that the Minkowski vacuum state is in fact a thermal state when restricted to $R^+$.

Let us introduce coordinates $(\eta, \xi, x^2,\ldots,x^{D-1})$ that are suited to $R^+$.
The coordinates $\eta$ and $\xi$ are defined via
\begin{align}
x^0 &=: \xi \sinh \eta \,,\\
x^1 &=: \xi \cosh \eta \,,\\
\eta \in (-\infty,&\infty),\quad \xi \in (0,\infty)\, .
\end{align}
The boundary of $R^+$,  given by $x^1 = \vert x^0 \vert$,  maps to $\xi = 0$.
The line element (\ref{minkds2}) in these coordinates reads
\begin{align}
\d s^2 = - \xi^2 \d\eta^2 + \d \xi^2 + (\d x^2)^2 + \ldots + (\d x^{D-1})^2 \, .
\label{ds2rindler}
\end{align}
Note that $\partial_\eta$ is a Killing vector,  and is in fact equal to the boost generator $x^0 \partial_1 + x^1 \partial_0$.

The curve with fixed $(\xi, x^2,\ldots,x^{D-1})$,  that is,  with only the $\eta$ coordinate varying,  is a timelike trajectory with constant proper acceleration,  see figure \ref{figrindler}.
This can be verified by representing the $(x^0,x^1)$ part of the trajectory with proper time coordinate $\lambda$,  that is,  via the parameterization $\left(\xi_0 \sinh \frac{\lambda}{\xi_0},\xi_0 \cosh \frac{\lambda}{\xi_0}\right)$. 
We can now see that the norm of the acceleration vector $\frac{\d^2 x^\mu}{\d \lambda^2}$ is constant and is equal to $\frac{1}{\xi_0}$.
In fact,  since the Rindler wedge is translationally invariant in the $(x^2,\ldots,x^{D-1})$ directions,  we can say that we have an infinitely big flat slab,  placed at some fixed $\xi_0$ that is constantly accelerating perpendicular to itself.
Invoking the principle of equivalence,  this is a model for an infinitely big and flat planet with a uniform gravitational acceleration \cite{Sciama:1981hr,  Takagi:1986kn}.

Consider a CFT on Minkowski space and a scalar operator $\phi$ in this CFT with dimension $\Delta$.
We place the CFT in the Minkowski vacuum state.
For ease of notation,  we also denote the Minkowski coordinates by $(t,x,\vec{x})$ where $\vec{x}$ is a $(D-2)$-dimensional vector.
The two-point function of $\phi$ is given by\footnote{We consider a Wightman function and there is an implicit $\i \epsilon$ prescription in the formula below.
The $\i \epsilon$ is not relevant for our considerations here,  so we omit it for the sake of clarity.}
\begin{align}
\langle 
\phi(t_1,x_1,\vec{x}_1) \,
\phi(t_2,x_2,\vec{x}_2) 
\rangle = \left[ -(t_1 - t_2)^2 + (x_1 - x_2)^2 + (\vec{x}_1 - \vec{x}_2)^2  \right]^{-\Delta}\,.
\label{correlator}
\end{align}
We now consider the situation where $(t_1,x_1,\vec{x}_1) \in R^+$ and also $(t_2,x_2,\vec{x}_2) \in R^+$,  such that the two points have the same $\xi$ and $\vec{x}$ coordinates,  see figure \ref{figrindler}.
In other words,  if there was no time separation between them,  the two operators would be identical.\footnote{We consider this case for simplicity,  since from the perspective of our discussion of the KMS property,  two field operators at different spatial locations are different operators.}

Let us now calculate the correlator in terms of the $\eta, \xi$ coordinates.
The squared separation between the two points appearing in the correlator (\ref{correlator}) becomes
\begin{align}
-(t_1 - t_2)^2 + (x_1 - x_2)^2
&= - \xi^2 (\sinh \eta_1 - \sinh \eta_2)^2
+ \xi^2 (\cosh \eta_1 - \cosh \eta_2)^2 \\
&= 2\xi^2 \left( 
1 - \cosh (\eta_1 - \eta_2)
\right)
\end{align}
Without loss of generality,  we let $\eta_2 = 0$.
Clearly since $\cosh(z) = \cosh(z')$ when $z + z' = -2\pi \i$,  we conclude that the correlator\footnote{We emphasize that this correlator is computed in the Minkowski vacuum.} $\langle  \phi(\eta) \,\phi(0) \rangle$ obeys the KMS property (\ref{kms}) with respect to the generator of $\eta$-translations and with $\beta = 2\pi$.
It is clear from the structure of the argument that this result does not depend on us having a CFT or considering a scalar operator;  we only used the expression of the Minkowski-invariant interval in Rindler coordinates.

The operator that generates the $\eta$-translations is the boost operator in Minkowski space,  but its interpretation in the wedge $R^+$ in the coordinate system (\ref{ds2rindler}) is that of a Hamiltonian since $\eta$ is the natural time coordinate in $R^+$.
Thus we arrive at the conclusion that Minkowski vacuum state (i.e.  the ground state of the operator that generates translations in $x^0$) is a thermal state from the point of the view of the Rindler Hamiltonian with $\beta = 2\pi$.

Usually,  this derivation is presented using a path integral in Euclidean space that prepares the Minkowski vacuum.
That derivation is,  of course,  physically correct and provides some very useful pieces of intuition.
We have chosen to present things using purely Lorentzian correlators since,  for a black hole formed from the collapse of a star,  the relevant state is not that of global thermal equilibrium.
The strategy of \cite{Fredenhagen:1989kr} to derive Hawking radiation is to show that two-point functions computed at late times in the black hole spacetime resulting from the collapse of a star obey the KMS property.

The quick way to figure out the temperature is to consider the line element (\ref{ds2rindler}) with purely imaginary $\eta$.
The proper framework to think about this is that the Minkowski manifold under consideration is a half-dimensional subspace of a larger manifold with complex dimension $D$.
If the quantities under study are analytic in this larger manifold,  one can take different half-dimensional sheets.
In particular,  let us take $\eta = - \i \theta$ with $\theta \in \mathbb{R}$.
The line element (\ref{ds2rindler}) becomes
\begin{align}
\d s^2 = \xi^2 \d\theta^2 + d\xi^2 + \d \vec{x}^2 \,.
\end{align}
Now the $(\theta, \xi)$ part of this metric is just a 2D Euclidean plane in polar coordinates.
If we demand the absence of a conical singularity at $\xi = 0$ in this geometry,  we must identify $\theta$ with periodicity $2\pi$.
The absence of a conical singularity is just the statement that the Minkowski vacuum state on $\mathbb{R}^{D-1}$ can be prepared using a path integral on half of $\mathbb{R}^{D+1}$.

\subsection{Temperature of an eternal Schwarzschild black hole}
The above discussion of Rindler space vis-a-vis Minkowski space naturally generalizes to the eternal Schwarzschild black hole.
Consider,  for concreteness,  the line element of a 4d Schwarzschild black hole in flat space
\begin{align}
\d s^2 &= - f(r) \d t^2 + \frac{\d r^2}{f(r)} + r^2 \d \Omega_2^2 \, , \label{ds2sch}\\
f(r) &= 1 - \frac{r_+}{r}\,,\quad r_+ > 0\, .
\end{align}
Here $r_+$ is the Schwarzschild radius,  or the location of the horizon,  and $\d \Omega_2^2$ is the standard metric on a 2-sphere.
Note that $f(r_+)=0$ and $f'(r_+) \neq 0$.

We consider a quantum field propagating in this eternal Schwarzschild black hole.
Since there is no global timelike Killing vector,  we cannot identify a natural analog of the Minkowski vacuum state,  which is the state with lowest eigenvalue of the quantum operator that generates translations along that timelike Killing vector.

However, the metric (\ref{ds2sch}) has a natural Euclidean continuation.
As in the Rindler case,  let $t = - \i \theta$ with $\theta \in \mathbb{R}$.
\begin{align}
\d s^2 &= f(r) \d \theta^2 + \frac{\d r^2}{f(r)} + r^2 \d \Omega_2^2 \, , \label{ds2scheuc}\\
\d s^2 &\approx f'(r_+) (r-r_+)  \,\d \theta^2 + \frac{\d r^2}{f'(r_+)(r-r_+)} + r_+^2 \d\Omega_2^2 \,,
\end{align}
where in the second line we have written the near-$r_+$ behavior of the metric.
If we define a new coordinate $\rho := \frac{2}{\sqrt{f'(r_+)}} \sqrt{r-r_+}$,  then we can write the near-horizon metric as
\begin{align}
\d s^2 \approx \frac{f'(r_+)^2}{4} \, \rho^2\,\d\theta^2  + \d\rho^2 +  r_+^2 \d\Omega_2^2 \,.
\end{align}
Thus,  in the $(\theta, \rho)$ section of the metric,  we again get a 2D plane in polar coordinates.
Demanding the absence of a conical singularity leads us to demand that the $\theta$ coordinate should be periodic with period
\begin{align}
\beta = 2\pi \cdot \frac{2}{f'(r_+)} = \frac{4\pi}{f'(r_+)}\, .
\end{align}

The Hartle-Hawking state of quantum fields in the eternal black hole geometry is the unique state such that correlators of operators in one of the exterior regions of the eternal Schwarzschild geometry are analytic in the strip $-\beta < \text{Im}(t) < 0$.
They also satisfy the KMS condition with respect to the generator of $t$-translations with $\beta = \frac{4\pi}{f'(r_+)}$.
Thus,  the Hartle-Hawking state is a nonzero temperature state when restricted to one of the outside regions.
Of course,  it is a pure state when we consider it as a state on the full Cauchy slice of the eternal Schwarzschild black hole geometry,  which includes both the left and the right exterior regions.

\subsection{A spectral paradox in the eternal black hole spacetime}
The most important difference between the Schwarzschild and the Rindler cases is that the Schwarzschild black hole is supposed to describe a finite temperature state of some putative quantum system with a \emph{finite} entropy.
The value of this finite entropy is given by the Bekenstein-Hawking area formula (\ref{sbek}).
Since the area of the Rindler horizon is infinite (this is the area spanned by coordinates $(x^2,\ldots, x^{D-1})$),  the value of the entropy,  even if a meaningful notion existed,  would be infinite.
For the black hole case,  the argument is most strong when the asymptotic spacetime is negatively curved anti-de Sitter space \cite{Maldacena:1997re, Witten:1998zw}.
There,  the black hole describes a finite temperature state of quarks and gluons in some gauge theory.

Consider now the operator $H$ that generates $t$-translations,  where $t$ is the Schwarzschild time coordinate in an eternal AdS Schwarzschild black hole.
We will describe a paradox by first presenting an argument that the spectrum of $H$ is discrete and then presenting an argument that the spectrum of $H$ is continuous.
The argument we present is a disguised version of \cite{Maldacena:2001kr}.

The argument that the spectrum of $H$ is discrete is simply that near the boundary of anti-de Sitter space,  the $t$-translations are the time translations of the boundary quantum system.
The Hamiltonian of the boundary gauge theory with finite $N$ and when placed on a spatial manifold of finite volume has a discrete spectrum.

The argument that the spectrum of $H$ is continuous can be phrased  mathematically by saying that,  in the bulk,  $H$ is proportional to the logarithm of the density matrix of bulk quantum fields restricted to,  say,  the right exterior region.
This is also known as the modular Hamiltonian.
The algebra of observables restricted to the right exterior region forms a so-called type-III$_1$ von Neumann algebra,  and the associated modular Hamiltonian always has a continuous spectrum \cite{Witten:2018zxz}.
Physically,  the continuous spectrum can also be seen by studying the time-separated two-point function of a simple boundary operator,  and noting that the presence of the black hole horizon implies an exponential decay of the two-point function in time.
A correlation function cannot decay to zero in a quantum system with finite entropy \cite{Maldacena:2001kr}.

Of course,  there is a loophole in this paradox.
The loophole is simply that the argument that shows that the spectrum of $H$ is continuous relies on a single classical bulk geometry,  and is thus only valid in the strict $N = \infty$ limit.
The boundary quantum system only has finite entropy at finite $N$.
When $N$ is finite,  the bulk geometry undergoes fluctuations,  and the Schwarzschild black hold may not be the only relevant geometry.
Indeed,  wormhole contributions to time-separated boundary two-point functions have been recently shown to give rise to small contributions,  suppressed by exponential of the entropy,  but which \emph{grow} in time \cite{Saad:2018bqo,Saad:2019pqd}.
This example teaches us a valuable lesson,  which is that we should never directly compare results obtained using the classical black hole geometry to microscopic predictions when those microscopic results rely on the finer aspects of the structure of the boundary quantum theory.

\subsection{Black holes formed from collapse}
It is harder to analyze a black hole formed from the collapse of a star as compared to an eternal black hole.
The main idea is that after the transients settle down,  the late time geometry is indeed described by the Schwarzschild metric,  but there is no past horizon.
The eternal black hole is kept in equilibrium by a balance between the radiation emitted by the black hole and the incoming radiation impinging on the black hole from past null infinity.
In the absence of a past horizon,   there is nothing to compensate the radiation emitted by the black hole,  and hence the black hole ``evaporates" \cite{Hawking:1974rv, Hawking:1975vcx}.

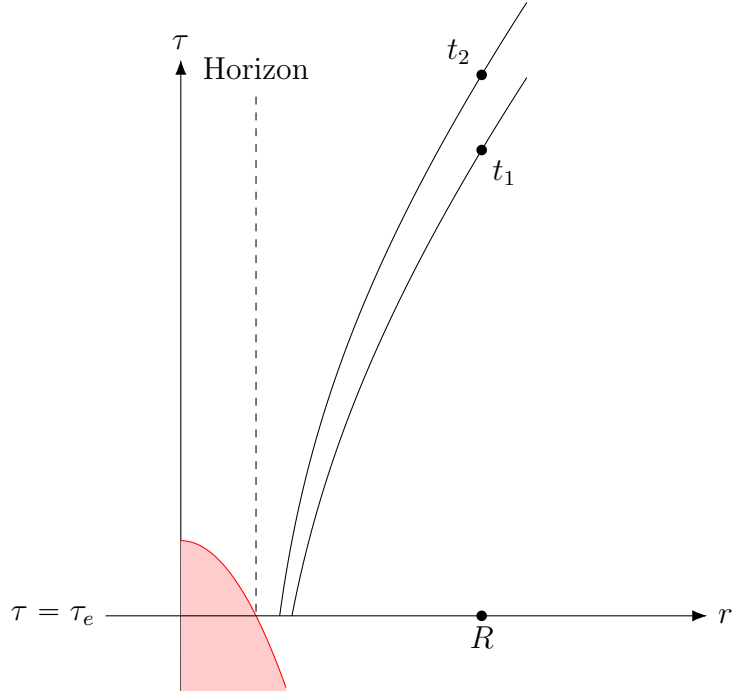
\begin{figure}[t!]
\centering
\begin{tikzpicture}

    \draw[-{Latex[length=2mm]}] (0, -1) -- (0, 7.4) node[above] {$\tau$};
    
    \draw[dashed] (1, -1) -- (1, 7) node[above] {Horizon};

	\draw[domain=1.477:4.6, samples=100, smooth, variable=\r] 
        plot ({\r},{\r + 2*ln(\r-1)});
     \fill (4,6.19722) circle (2pt) node[below right] {$t_1$};
     
     \draw[domain=1.314:4.6, samples=100, smooth, variable=\r] 
        plot ({\r},{\r + 2*ln(\r-1) + 1});
   	\fill (4,7.19722) circle (2pt) node[above left] {$t_2$};

        \fill[red!20] 
        plot[domain=0:1.4, samples=100, smooth, variable=\r] ({\r}, {1 - \r^2}) --
        plot[domain=1.4:0, samples=100, smooth, variable=\r] ({\r}, {-1}) -- cycle;
        
      \draw[domain=0:1.4,samples=100,smooth,variable=\r,red]
       plot({\r},1-\r^2);
       
           \draw[-{Latex[length=2mm]}] (-1, 0) -- (7, 0) node[right] {$r$};

    \node at (-1.7, 0) {$\tau = \tau_e$};
    \fill (4,0) circle (2pt) node[below] {$R$};

\end{tikzpicture}
\caption{The geometry of a black hole formed from collapse.
The red shaded region represents the collapsing star and the dotted vertical line is the horizon.  We are interested in computing the two-point function of a free scalar field,  with the two operators inserted at radial location $R$ and at late times $t_1$ and $t_2$.
The outgoing light rays passing through these two points are shown,  they emanate from locations very close to the horizon,  and thus from locations that are close to each other.
This two-point function is controlled by a short-distance two-point function on the $\tau = \tau_e$ slice which is universal for smooth states in a field theory with a known UV fixed point.}
\label{figcollapse}
\end{figure}

We will present a short account of the derivation of Hawking radiation given by Fredenhagen and Haag \cite{Fredenhagen:1989kr},  leaving most of the details to the original reference.
Say we have a free massless scalar field that is minimally coupled to the background metric of the spacetime.
The main idea is to compute a time-separated two-point function of the scalar field at a fixed radial location far away from the black hole horizon,  with both time points being sufficiently late compared to the moment when the star crosses its event horizon.

The main idea is that two-point function of a scalar field also obeys the Klein-Gordon wave-equation.
So,  given the hyperbolic nature of the wave-equation,  the two-point function at late times can,  very roughly,  be thought as being transported along light rays. 
Now,  light rays that arrive at a faraway location at late times all emanate from a region very close to the black hole horizon.

Let us see this in more detail.
Let $r_*$ be the tortoise coordinate defined via
\begin{align}
r_* := r + r_+ \log \left( \frac{r}{r_+} -1 \right)\, .
\end{align}
Note that $r_* \in (-\infty, \infty)$ as $r \in (r_+,\infty)$,  and also that $\d r_* = \frac{\d r}{f(r)}$.
Near infinity,  we have $r_* \approx r$ upto a logarithmic correction,  while near the horizon it is the logarithmic term that dominates 
$r_* \approx r_+ \log  \left( \frac{r}{r_+} -1 \right)$.

Define $u$ and $v$ coordinates via the relations
\begin{align}
v &:= t + r_*\,,\quad 
u := t - r_*\, .
\end{align}
Since the $(t,r)$ part of the Schwarzschild metric may be written equivalently  as $ -f \,  \d u \d v$,  the curves of constant $v$ or constant $u$ represent light rays.
The curves with constant $v$ are light rays that go in to the black hole.
The curves with constant $u$ are light rays that go out towards infinity.
Now define a new time coordinate $\tau$ via
\begin{align}
\tau := v - r = t + r_* - r\, .
\end{align}
The idea is the while the $(t,r)$ coordinates only cover the exterior region,  the coordinates $(v,r)$ or $(\tau,r)$ also cover the region behind the horizon.
Thus the slices $\tau=\text{constant}$ can serve as nice slices that cross the black hole horizon smoothly,  see figure \ref{figcollapse}.
Note also that $\tau \approx t$ far away from the black hole.

Imagine a light ray that arrives at a point $(t,R)$ where $t$ and $R$ are both sufficiently large.
We can imagine a detector placed at the location $R$ that measures Hawking radiation arriving at that location.
We know that,  a long time in the past,  say at some small value of $\tau$,  this light ray must have been very close to the horizon.
We want to work out the precise relationship. 
Recalling that outgoing light rays have constant value of the $u$ coordinate,  
and using $u = \tau + r - 2 r_*$,  we can equate the values of $u$ at the final and the initial points on the light ray to get
\begin{align}
t - R \approx \tau_e + r_+ - 2 \left( r_+ + r_+ \log \left(\frac{r}{r_+} - 1\right) \right)\, .
\end{align}
We have substituted $r \approx r_+$ everywhere except inside the logarithm.
On the left hand side,  we imagine keeping $R$ fixed and taking larger and larger value of $t$.
The dominant contribution from the right hand side is the logarithmic dependence of the tortoise coordinate near the horizon, and thus we get the relationship
\begin{align}
r - r_+ \approx r_+ \exp \left( - \frac{t-R}{2r_+} \right) \times \text{(an order one number)} \, .
\label{rminusrplus}
\end{align}
This equation tells us that for a light ray to arrive at later and later times at some fixed location $R$ far away from the black hole,  the light ray must have been emitted closer and closer to the horizon: $r - r_+ \sim r_+ \exp (-\frac{t}{2r_+})$,  where $r$ is the location from where the ray started at the early time $\tau = \tau_e$;  see figure \ref{figcollapse}.

This means that if we have two points $(t_1, R)$ and $(t_2,R)$ at which we are trying to measure the correlation function of a free quantum field,  that correlator will be controlled,  via a hyperbolic wave equation,  by the two-point correlator at points $(\tau_e, r_1)$ and $(\tau_e, r_2)$ where $r_1$ and $r_2$ are both very close to the horizon,  with $r_1 - r_+ \sim r_+ \exp (-\frac{t_1}{2r_+})$ and $r_2 - r_+ \sim r_+ \exp (-\frac{t_2}{2r_+})$.

Now comes another crucial piece of input.
We believe that the state of quantum fields in the vicinity of the horizon must be smooth,  since there are no curvature invariants getting large at the horizon.
Any state that is smooth has a universal behavior of correlation functions in the short-distance limit: it is simply that the short-distance correlation function should be the same as that in the vacuum state \cite{Haag:1984xa,Fredenhagen:1986jg, Witten:2021jzq}.
Violations of this condition are possible if the state under consideration is singular,  or,  in other words,  if it contains energy densities at arbitrarily short length scales.
We assume this is not the case.\footnote{This assumption was challenged in the light of the Mathur-AMPS paradox \cite{Mathur:2009hf, Almheiri:2012rt},  or what has come to be knows as the ``firewall problem."}

Since the vacuum two-point function of a free scalar field in four spacetime dimensions is $d^{-2}$, where $d$ is the distance between the two points,  we conclude that the late time two-point function of interest in the black hole geometry is proportional to $(r_1 - r_2)^{-2}$.
Eq.(\ref{rminusrplus}) then tells us that the variables $t_1$ and $t_2$ enter via the exponentials $\exp (-\frac{t_1}{2r_+})$ and $\exp (-\frac{t_2}{2r_+})$.
This exponential dependence on time implies the KMS condition with the required value of $\beta = 2\pi \cdot 2 r_+ = 4\pi r_+$ (\ref{thawk}).
We refer the reader to \cite{Fredenhagen:1989kr} for further details.
  
It is possible to approximately calculate the expectation value of the stress tensor in a model of the background geometry and the quantum state and to compute how this stress tensor backreacts and modifies the geometry \cite{Visser:1997sd, Abdolrahimi:2016emo}.
While the calculations are very hard in four dimensions,  it is possible to make models in two-dimensions and analytically calculate the Hawking flux \cite{Callan:1992rs,  Strominger:1994tn}.
It is important to note a crucial difference between the eternal and the evaporating black holes: The off-diagonal component of the stress-tensor $T^{0i}$ which measures the energy flux is zero in the eternal black hole geometry,  but it is nonzero in the evaporating black hole.
In the eternal black hole,  the flux has positive and negative contributions that cancel,  as expected in a thermal equilibrium state.

\section{Page curve for an eternal black hole}
\subsection{Setup}
Since mathematical treatment of an evaporating black hole is harder,  we focus on the case of an eternal black hole.
A version of the entropy paradox and the Page curve exists even in this geometry \cite{Mathur:2014dia,Almheiri:2019yqk}.
We now describe this paradox.

Imagine that we have a stable eternal black hole in anti-de Sitter space.
This geometry describes the thermofield double state of the boundary CFT \cite{Maldacena:2001kr}.
The thermofield double state of any quantum system with energy eigenvalues $E_i$ and corresponding eigenvectors $\ket{E_i}$ is defined as the following state on two-copies of the system:
\begin{align}
\ket{\Psi^{\text{tfd}}_{\beta}} := \frac{1}{\sqrt{Z(\beta)}} \sum_{i} e^{-\frac{\beta E_i}{2}}
\ket{E_i} \ket{E_i}\, .
\label{psitfd}
\end{align}
This state has the property that if we trace out one of the copies,  the reduced density matrix on the other copy is the equilibrium thermal density matrix.

Now,  a stable AdS black hole does not evaporate,  but one can model the evaporation as follows.
We couple the left copy of the boundary quantum system to one heat bath,  and the right copy of the boundary quantum system to another heat bath.
Here,  by coupling we mean that the Hamiltonian contains terms that couple the boundary quantum system and the heat bath.
We imagine the number of degrees of freedom in the heat bath to be much larger than the entropy of the boundary quantum system at the chosen temperature.
We put the combined system into a left-right thermofield double state.
Note that the left and right systems are not coupled by any terms in the Hamiltonian,  but the state is left-right entangled.
Now,  we evolve the combined system (made of four total sub-parts) by the Hamiltonian
\begin{align}
H = H_{\text{left}} + H_\text{right}\, .
\end{align}
It is easy to see that the state (\ref{psitfd}) evolves non-trivially when we apply the operator $e^{-\i H t}$.
This simple model of time evolution in what would otherwise have been an equilibrium state was studied in \cite{Hartman:2013qma}.

Let $Q_1$ and $Q_2$ denote the left and right copy of the boundary quantum system that is holographically dual to the theory of quantum gravity in AdS.
Let $R_1$ and $R_2$ denote the two heat reservoirs.
The reservoir $R_1$ is coupled to $Q_1$,  and the reservoir $R_2$ is coupled to $Q_2$.
Then,  with the initial state and the time-evolution operators described above, we ask the question: What is the von Neumann entropy of the system $R_1 \cup R_2$ as function of time?

Since the initial state is not a stationary state with respect to the time-evolution operator,  we expect that, at early times, the von Neumann entropy will increase linearly with time \cite{Calabrese:2005in,Hartman:2013qma},  as it does in situations involving a so-called ``quantum quench."
However,  this increase cannot continue forever.
The combined system $Q_1R_1Q_2R_2$ is in a pure state and so the entropy of $R_1 \cup R_2$ is the same as the entropy of $Q_1 \cup Q_2$.
However,  the entropy of $Q_1 \cup Q_2$ is bounded above by $2 S(\beta)$, where $S(\beta)$ is the entropy of the thermal density matrix of either of the quantum systems $Q_1$ and $Q_2$.
In the gravity description,  this upper bound will be $2 S_\text{BH}$,  where $S_{\text{BH}}$ is the Bekenstein-Hawking area-entropy of a black hole at inverse temperature $\beta$.
Thus,  we expect that the entropy will increase linearly until it reaches the upper bound, and which point,  there should be a phase transition in the behavior of the entropy and the entropy curve should then become flat. 

The gravity description of this system is as follows.
The system $Q_1 \cup Q_2$ is dual to a two-sided AdS black hole.
To model the heat reservoirs,  we simply let the quantum fields in the black hole spacetime leak out from the boundary of AdS space,  rather than let them reflect back in.
We can imagine that the energy leaks from AdS space into (and back from) a flat space region which does not have gravity,  and which represents the reservoir into which the Hawking radiation escapes. 
So the question about the entropy of Hawking radiation becomes a question about the entropy of these reservoir systems.
In the absence of an incoming flux,  this will lead to continuous loss of energy from the black hole,  and the black hole will shrink \cite{Almheiri:2019psf}.
The introduction of this energy loss mechanism comes at the cost of losing some rigor in applying the AdS/CFT correspondence,  and also about the nature of gravity in the presence of such boundary conditions \cite{Raju:2020smc},  but the issues that arise are strongly believed to be red herrings for the purposes of the entropy computations that we will discuss.
The only purpose that the reservoirs are serving is to allow us to split the total subsystem into a black hole region and a radiation region,  so we can define and compute what we mean by the entropy of Hawking radiation.

To summarize the above discussion, we have an eternal two-sided black hole geometry in AdS,  and the two boundaries are coupled to flat space reservoirs with no gravity.
The full system is in the thermofield double state and we evolve the system forwards on both sides (as opposed to Killing time flow,  for which the time evolution of the two sides are oppositely directed).
Now,  the information paradox in this setting is that,  in the absence of an intervening mechanism, semi-classical gravity will predict that the linear-in-time increase of the entropy of the reservoirs will continue forever.
However,  we know that the entropy must saturate.\footnote{These is no debate that the Page curve exists for realistic quantum systems.  See \cite{Saha:2024ims} for explicit computations which also defines a coarse-grained entropy that continues to increase after the Page time. }
In gravity,  we will see that the quantum extremal surface (QES) prescription \cite{Engelhardt:2014gca},  with an allowance for disconnected entanglement wedges \cite{Penington:2019npb,Almheiri:2019psf,Almheiri:2019hni},  reproduces the Page curve.
The nontrivial QES stems from wormhole contributions to path integral replica computations of the entropy.

In this section,  we describe the computations of \cite{Almheiri:2019yqk} that show this in detail.
However,  before that,  in the next two subsections,  we introduce some background material.
First,  we present  some results for the entanglement entropy in 2d CFTs,  and then we introduce JT gravity (coupled to matter and with heat reservoirs),  the gravity theory for in which the QES computations will be performed.

We end this prelude by briefly introducing the QES formula.
We want to compute the entropy of some boundary spatial subregion $A$,  with the full boundary theory being in a specified quantum state.
The holographic computation of the entropy of $A$ proceeds by considering candidate co-dimension one regions $R$ in the bulk geometry that is dual to the boundary state.
The only restriction is that the boundary of $R$ should be homologous to $A$; this allows $R$ to have disconnected components.
Then one constructs the generalized entropy function
\begin{align}
\sgen(R) := \frac{\text{Area}(\partial R)}{4G} + S_\text{bulk}(R)\, ,
\label{sgen}
\end{align}
where $S_\text{bulk}(R)$ is the entropy of matter fields in the bulk region $R$ when we trace out the complement of $R$ in the bulk.
The result \cite{Engelhardt:2014gca,Dong:2017xht} is that the entropy of $A$ is given by $\sgen$ evaluated on the region $R_\star$ which minimizes the generalized entropy.\footnote{The ``minimization" needs some qualifiers which are not important for our purposes.}
The region $R_\star$ is the entanglement wedge of $A$: all bulk EFT operators that are localized in $R_\star$ can be represented only in the boundary region $A$ \cite{Dong:2016eik,Cotler:2017erl,Chen:2019gbt,Penington:2019kki}.

\subsection{Some entanglement entropy formulas in two dimensions}
\label{sec:cftentropy}

Consider a 2d CFT in its ground state on an infinite spatial line.
We take the spacetime metric to be
\begin{align}
\d s^2 &= - \d t^2 + \d x^2 = - \d x^+ \,\d x^- \,,\quad \text{where} \\
x^+ &:= t + x \quad  \text{ and } x^- := t - x \, .
\end{align}
Consider the reduced state on an interval $PQ$ of (proper) length $\ell$.
It is well-known result that \cite{Holzhey:1994we,  Calabrese:2004eu,Calabrese:2009qy} that the von Neumann entropy of this reduced state $S(\ell)$ is given by
\begin{align}
S(\ell) = \frac{c}{3} \log \frac{\ell}{\varepsilon}\, ,
\label{svacl}
\end{align}
where $c$ is the central charge of the theory and $\varepsilon$ is the UV cutoff with dimensions of length.
If the line segment $PQ$ is such that $P$ and $Q$ do not have the same value of the time coordinate,  then the entropy formula is the covariant version of the above (note that the interval $PQ$ has to be spacelike)
\begin{align}
S(PQ) &= \frac{c}{6} \log (-(x^+(Q) - x^+(P))(x^-(Q) - x^-(P)))\, ,  \label{spqcov} \\
&= \frac{c}{6} \log \vert \Delta x^+ \vert + \frac{c}{6} \log \vert \Delta x^- \vert \, , \label{spqleftright}
\end{align}
where we have dropped the cutoff $\varepsilon$ in this formula.
Note that the argument of the logarithm in (\ref{spqcov}) is positive since the interval $PQ$ is spacelike.

Next,  we would like to have some formulas for the entanglement entropy when the background spacetime is not flat but instead has some curvature,  with the metric related to the Minkowski line element by a Weyl factor:
\begin{align}
\d s^2 := \Omega(t,x)^{-2}\, (-\d t^2 + \d x^2)\, .
\label{omegadef}
\end{align}
We consider the state in this curved space which is the Weyl transform,  by the function $\Omega$,   of the ground state in the standard flat metric $-\d t^2 + \d x^2$.
The Weyl factor contributes to the entropy as
\begin{align}
S(PQ) = \frac{c}{6} \log (-(x^+(Q) - x^+(P))(x^-(Q) - x^-(P))) - \frac{c}{6} \log \Omega(P) - \frac{c}{6} \log \Omega(Q)\, .
\label{spqomega}
\end{align}
The final two terms are the extra pieces that we obtain in curved space.
These contributions can be deduced by noting that the UV cutoff on the coordinates depends on $\Omega$ in such a way that the UV cutoff on the proper length is some fixed number.
These contributions can also be seen as arising due to the Weyl transform of the twist operators whose correlation functions compute the Renyi entropies \cite{Calabrese:2004eu}.
Note that these extra contributions depend locally on the end-points of the interval.

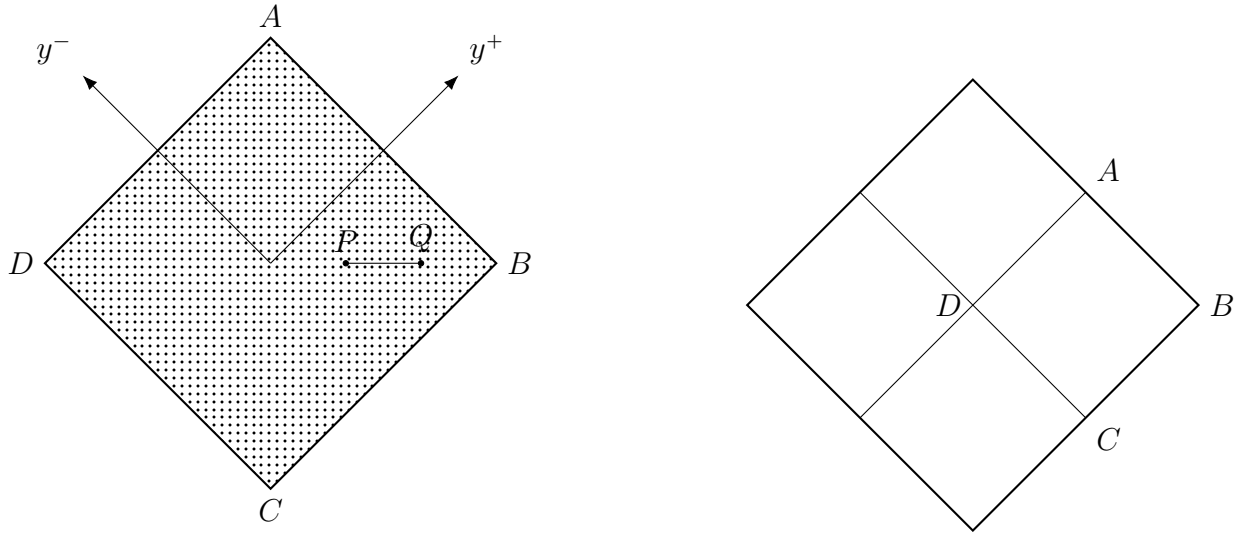
\begin{figure}
\centering
\begin{tikzpicture}

	\fill[pattern=dots] 
        (0,3) -- (3,0) -- (0,-3) -- (-3,0) -- cycle; 
    \draw[thick] 
        (0,3) -- (3,0) -- (0,-3) -- (-3,0) -- cycle;

    \draw[-{Latex[length=2mm]}] (0,0) -- (2.5,2.5) node[anchor=south west] {$y^+$};
    \draw[-{Latex[length=2mm]}] (0,0) -- (-2.5,2.5) node[anchor=south east] {$y^-$};

    \filldraw (1,0) circle (1pt) node[anchor=south] {$P$};
    \filldraw (2,0) circle (1pt) node[anchor=south] {$Q$};
    \draw[-] (1,0) -- (2,0);

	\node[above] at (0,3) {$A$};
	\node[right] at (3,0) {$B$};
	\node[below] at (0,-3) {$C$};
	\node[left] at (-3,0) {$D$};
\end{tikzpicture}
\hfill
\begin{tikzpicture}

	\draw[thick] 
        (0,3) -- (3,0) -- (0,-3) -- (-3,0) -- cycle;  
     \draw (1.5,1.5) -- (-1.5,-1.5);   
     \draw (-1.5,1.5) -- (1.5,-1.5);   
     \node[above right] at (1.5,1.5) {$A$};
     \node[right] at (3,0) {$B$};
     \node[below right] at (1.5,-1.5) {$C$};
     \node[left] at (0,0) {$D$};
\end{tikzpicture}

\caption{The left figure shows the Penrose diagram $ABCD$ of a 2d Minkowski space $M_1$ in which a 2d CFT at nonzero temperature is placed.  We have depicted this hot matter by the dots.  We would like to compute the von Neumann entropy of the spatial interval $PQ$.
In the right figure,  the Minkowski space $M_1$ gets embedded as the right Rindler wedge of a bigger Minkowski spacetime $M_2$.
The state in $M_2$ in the vacuum.
The entropy of interest can be computed using the vacuum formulas in $M_2$ and applying the necessary Weyl transformation.
}
\label{figthermalcft}
\end{figure}

Let us use these results to compute the entropy of an interval in the thermal state of a 2d CFT,  still on an infinite spatial line.
Let $- \d y^+ \,  \d y^-$ be the metric on Minkowski space $M_1$ where this 2d CFT in the thermal state lives.
Following the discussion of the Bisognano-Wichmann theorem,  we embed this thermal CFT on $M_1$ as the right Rindler wedge of a bigger Minkowski spacetime $M_2$,  see figure \ref{figthermalcft}. 
The state of the CFT on $M_2$ will be the vacuum state and thus we can use the entropy formulas discussed above.
We let $\beta = 2\pi$ for now;  we will restore $\beta$ in the final formula at the end.
Let
\begin{align}
x^+ := \exp (y^+) \,,\quad x^- := -\exp (y^-)\, .
\label{xyrelation}
\end{align}
Note that $M_1$,  which is covered by the range $y^+ \in (-\infty,\infty)$ and $y^- \in (-\infty,\infty)$ gets mapped to the region $x^+ \in (0,\infty)$ and $x^- \in (-\infty,0)$ inside $M_2$.
The full space $M_2$ is covered by $x^+ \in (-\infty,\infty)$ and $x^- \in (-\infty,\infty)$.
The metrics are related by
\begin{align}
-\d y^+ \, d y^- &= - \frac{\d x^+ \, d x^-}{-x^+ x^-}\, ,  \quad \text{with } x^+ > 0 \text{ and } x^- < 0\,, 
\end{align}
and thus comparing to (\ref{omegadef}),  we have
\begin{align}
\Omega(x^+,x^-) &= (-x^+ x^-)^\frac{1}{2}\, .
\end{align}
Let us consider an interval $PQ$ (see left panel of figure \ref{figthermalcft}) with its various coordinates specified as follows
\begin{align}
\begin{tabular}{c|c|c}
& $P$ & $Q$ \\
\hline
$(t,y)$ & $(0,a)$  & $(0,b)$ \\
$(y^+,y^-)$ & $(a,-a)$ & $(b,-b)$ \\
$(x^+,x^-)$ & $(e^a,-e^a)$ & $(e^b,-e^b)$ \\
\end{tabular}
\end{align}
Here $(t,y)$ coordinates are related to $(y^+,y^-)$ by $y^+ = t + y$  and $y^- = t - y$.
So we simply have an interval of length $b - a$ (we take $b>a$) on an equal time slice.
We compute the entropy using (\ref{spqomega}):
\begin{align}
S(PQ) &= 
\frac{c}{6} \log \vert \Delta x^+ \vert + 
\frac{c}{6} \log \vert \Delta x^- \vert
- \frac{c}{6}\log \Omega(P)
- \frac{c}{6}\log \Omega(Q) \\
&= 2\times \frac{c}{6} \log (e^b - e^a) - \frac{c}{6} \log(e^a) - \frac{c}{6} \log (e^b) \\
&= \frac{c}{3} \log \left( 
2 \sinh \frac{b-a}{2}
\right)\, .
\end{align}
Restoring $\beta$,  we find
\begin{align}
S(PQ) = \frac{c}{3} \log \left( \frac{\beta}{\pi} \sinh \frac{\pi \ell}{\beta} \right)\, .
\label{sthermal}
\end{align}
This is a well-known result \cite{Calabrese:2009qy}.
We can perform two sanity checks: (a) In the zero temperature limit,  $\beta \to \infty$ and we recover the vacuum result (\ref{svacl}),  (b) When $\ell \gg \beta$,  we get an entropy proportional to the length $\ell$,  which is simply a statement of the extensivity of thermal entropy.

\subsection{JT gravity}
In order to perform detailed concrete computations of the entropy of Hawking radiation,  one needs a simple model of gravity.
One such model is JT gravity,  which is a two-dimensional theory of dilaton gravity \cite{Jackiw:1984je,  Teitelboim:1983ux,  Maldacena:2016upp}.
In certain parameter regimes,  it is dual to a disordered quantum mechanical system of Majorana fermions,  known as the Sachdev-Ye-Kitaev (SYK) model \cite{Sachdev:1992fk,  Maldacena:2016hyu, Kitaev:2017awl}.
The details of the SYK model will not be relevant for us;  we just need to know that it can be thought of a holographic system dual to JT gravity.

The Euclidean action for JT gravity is
\begin{align}
    16 \pi G\, I_\text{Euc} = &- \phi_0 \left( \int \d^2x\, \sqrt{g} \, R
    + \int \d x \, \sqrt{\gamma} \, 2K 
    \right) \notag \\ 
    & - \left( \int \d^2x\, \sqrt{g} \, \phi (R+2)
    + \int \d x \, \sqrt{\gamma} \, 2 \phi \, (K-1) 
    \right) \\
    &= - \phi_0 \, \chi - \left( \int \d^2x\, \sqrt{g} \, \phi (R+2)
    + \int \d x \, \sqrt{\gamma} \, 2 \phi \, (K-1) 
    \right) \, .
\end{align}
Here $\chi$ is the Euler characteristic of the surface on which we are studying the theory.
The quantity $\phi_0$ serves as the genus counting parameter of this theory,  and supplies an additive constant to the entropy of black holes (so it represents an entropy that is present even at zero temperature).
The quantity $\phi_0 + \phi$ is the total dilaton,  and can be thought of as the area of a near-extremal higher dimensional black hole that has been dimensionally reduced to two dimensions.
The $-1$ in $K-1$ is a holographic counterterm.

The action needs to be supplemented with boundary conditions. 
They are \cite{Maldacena:2016upp}
\begin{align}
    g_{tt}\big\vert_\text{bdry} &= - \, \varepsilon^{-2} \,, \label{bc1}\\
    \phi \big\vert_\text{bdry} &= \phi_r  \, \varepsilon^{-1}\,. \label{bc2}
\end{align}
The constant $\phi_r$ has dimensions of length and is a dimensionful parameter of the theory.
This will set the length scales of various things in the theory.
Often we will encounter it in dimensionless combinations like $\phi_r/\beta$.
Also $\varepsilon$ is a small quantity that should be understood to be the smallest length scale in the problem (UV cutoff), just like when we compute things by using cutoffs near the boundary in AdS.

We would now like to write down the equations of motion for this theory and find some simple black hole solutions.
For that,  let us remind ourselves how the Ricci scalar transforms when we make a variation of the metric. 
It is useful to vary the inverse metric:
\begin{align}
    \delta R = R_{ab} \, \delta g^{ab} - \nabla_a\nabla_b \, \delta g^{ab} 
    + \nabla^c \nabla_c \, g_{ab}\delta g^{ab}\, .
\end{align}
Terms that look like total derivatives are important since the Ricci scalar multiplies the dilaton in the action.
We can now vary the fields and get the equations of motion.
\begin{align}
R + 2 &= 0 \,,\\
\left( \nabla_c \nabla^c \phi \right) \, g_{ab} - \nabla_a \nabla_b \phi - \phi \, g_{ab} &= 0\,.
    \label{dilatoneom}
\end{align}

We will couple this gravity theory to matter to get the physics of Hawking radiation.
The matter sector will be taken to be a CFT$_2$ with central charge $c$ that is minimally coupled to the metric,  and not at all to the dilaton.

\subsection{QES at zero temperature}

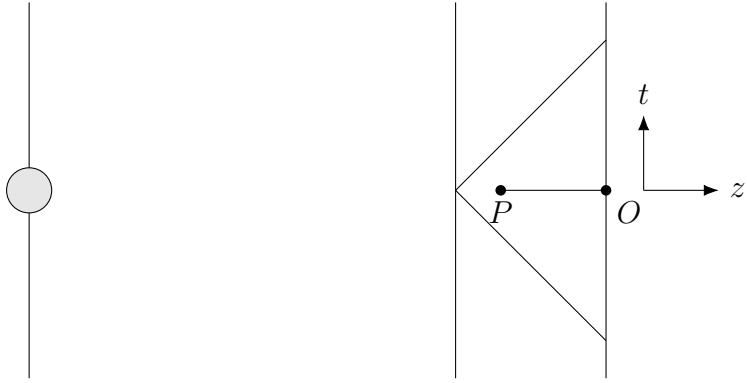
\begin{figure}
\centering
\begin{tikzpicture}

\draw (0,-2.5) -- (0, 2.5);
\filldraw[fill=gray!20] (0, 0) circle (0.3);

\end{tikzpicture}
\hspace{2in}
\begin{tikzpicture}
\draw (0, -2.5) -- (0, 2.5);
\draw (2, -2.5) -- (2, 2.5);

\draw[-] (0, 0) -- (2, 2);
\draw[-] (0, 0) -- (2, -2);

\draw[-{Latex[length=2mm]}] (2.5, 0) -- (3.5, 0) node[right] {$z$};
\draw[-{Latex[length=2mm]}] (2.5, 0) -- (2.5, 1) node[above] {$t$};

\draw (0.6,0) -- (2,0) node[below right] {$O$};
\node[below] at (0.6,0) {$P$};
\fill (0.6, 0) circle (2pt);
\fill (2.0, 0) circle (2pt);

\end{tikzpicture}

\caption{The left figure depicts an SYK quantum mechanical system,  with the time direction indicated.  The system is in its ground state. The dual geometry is the Poincar\'e patch of AdS$_2$.  We want to compute the entropy of the whole system.  The candidate QES is shown as the point $P$.  The entropy of bulk matter fields in the region $PO$ contributes to the generalized entropy function. }
\label{figzerot}
\end{figure}

The equation of motion $R +2 = 0$ says that the metric should have constant negative curvature.
The zero temperature solution is the metric of AdS$_2$ in the Poincar\'e patch
\begin{align}
    \d s^2 &= \frac{1}{z^2 } \left( - \d t^2 + \d z^2 \right) \, .
\end{align}
The boundary of AdS is at $z = 0$ but,  contrary to usual convention,  for later convenience, we take the range of the $z$ coordinate in AdS space to be $(-\infty, 0)$.
See figure \ref{figzerot}.

For $\phi$,  we will make the ansatz that it only depends on $z$.
Then the $zz$ component of the dilaton equation yields the most stringent equation which is first order,  namely $z \phi' + \phi = 0$.
This yields
\begin{align}
    \phi(z) &= \frac{A}{-z} \, .
\end{align}
Imposing the boundary conditions (\ref{bc1}) and (\ref{bc2}),  we see that the boundary curve should be the simple locus $z = - \varepsilon$ and we also must have $A = \phi_r$.

This field configuration should be thought of as describing the SYK model in its zero temperature state.

We want to apply the QES prescription \cite{Engelhardt:2014gca} to compute the entropy and entanglement wedge of the full boundary system.
Clearly,  we know that the entanglement wedge should be the full AdS geometry.
Let us see this from a computation.

Since the matter CFT has fully reflecting boundary conditions,  we do not get independent contributions to the entropy from the left and right movers as in (\ref{spqleftright});
we just get one contribution with coefficient $\frac{c}{6}$.
Let the candidate surface be a point $P$ with coordinates $(0,-a)$ with $a>0$.
The the generalized entropy function (\ref{sgen}) is
\begin{align}
    \sgen(a) &= \frac{\phi_r}{a} + \left( \frac{c}{6} \log (a) - \frac{c}{6} \log \Omega(P) \right)= \frac{\phi_r}{a}\,.
\end{align}
We have set $4G=1$ and used the fact that the value of the dilaton plays the role of area in JT gravity.\footnote{We have also dropped the constant $\phi_0$ since it is not important for the extremization.}
The second and third term in the middle formula canceled each other.
Clearly,  $\sgen(a)$ is extremized at $a_\star = \infty$.
This says that the QES is located at the Poincar\'e horizon and thus the entanglement wedge is all of AdS,  as expected.

\begin{figure}
\centering
\begin{tikzpicture}

\draw[-{Latex[length=2mm]}] (-1,-2) -- (-1, 2) node[above]{$t$};
\filldraw[fill=gray!20] (0, 0) circle (0.3);
\draw (0.3,0) -- (2.5,0);

\fill (0.7, 0) circle (2pt);

\node[below right] at (0.7, 0) {$\mathsf{O}$};

\end{tikzpicture}
\hspace{0.5in}
\begin{tikzpicture}

\draw (0, -2.5) -- (0, 2.5);
\draw (2, -2.5) -- (2, 2.5);

\draw[-] (0, 0) -- (2, 2);
\draw[-] (0, 0) -- (2, -2);
\draw[-] (2, 2) -- (4, 0);
\draw[-] (4, 0) -- (2, -2);


\draw (0.8,0) -- (2.4,0) node[below right] {$O$};
\node[below] at (0.8,0) {$P$};
\fill (0.8, 0) circle (2pt);
\fill (2.4, 0) circle (2pt);

\end{tikzpicture}
\hspace{0.5in}
\begin{tikzpicture}

\draw (0, -2.5) -- (0, 2.5);
\draw (2, -2.5) -- (2, 2.5);

\draw[-] (0, 0) -- (2, 2);
\draw[-] (0, 0) -- (2, -2);
\draw[-] (2, 2) -- (4, 0);
\draw[-] (4, 0) -- (2, -2);



\fill[gray!40] (2.4, 0) -- (3.2, 0.8) -- (4, 0) -- (3.2, -0.8) -- cycle;
\fill (2.4, 0) circle (2pt);
\node[below] at (2.4,0) {$O$};

\fill[gray!40] (0.6, 0) -- (0.3, 0.3) -- (0, 0) -- (0.3, -0.3) -- cycle;
\fill (0.6, 0) circle (2pt);
\node[below right] at (0.6,0) {$P_\star$};

\end{tikzpicture}

\caption{The left figure shows an SYK model coupled to a reservoir which is a semi-infinite wire supporting a 2d CFT.  The system is in its combined ground state.  We want to compute the entropy and entanglement wedge of a system consisting of the SYK model plus a portion of the wire contiguous to it,  upto the point labeled by $\mathsf{O}$.  
This is modeled in the gravity description (middle figure) on the right by attaching a non-gravitating flat space region to AdS$_2$.  The matter fields can now propagate freely across the interface.  The candidate QES is depicted by the point $P$.
The actual QES $P_\star$ for this bipartite decomposition of the system lies away from the Poincar\'e horizon.
This means that the entanglement wedge of the subsystem to the right of $\mathsf{O}$ in the left figure,  shown by the grey shaded region on the right,  is disconnected.
}
\label{figzerotcoupled}
\end{figure}
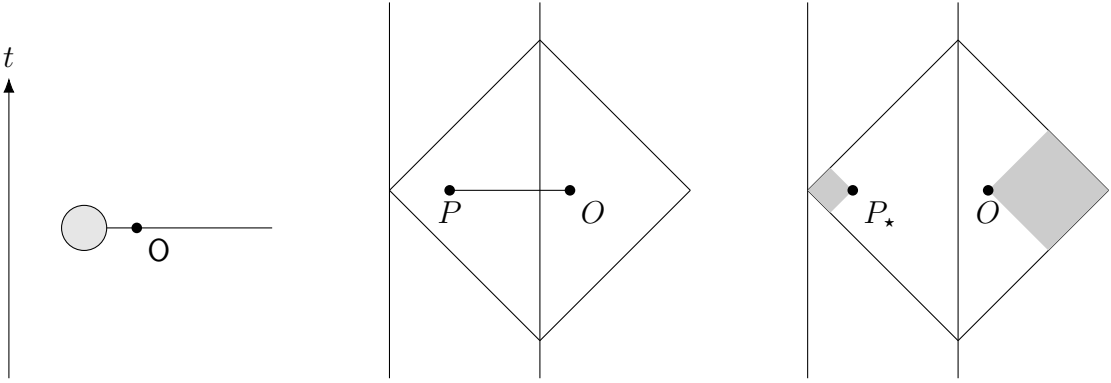

We now consider a different problem.
Let us couple the SYK system to a reservoir and place the combined system in its ground state.
The reservoir takes the form of a semi-infinite ``wire" and the SYK system is coupled to it at its endpoint.
We now want to compute the entropy and the entanglement wedge of the SYK model,  after tracing out the reservoir.
We can also keep a small contiguous part of the reservoir together with SYK and consider this as the subsystem of interest.
So,  we are interested in the entropy and entanglement wedge of the subsystem $[0,b]$ of this quantum mechanical system.
It is very important to emphasize that the entropy that is being computed is that of a subsystem of a non-gravitational system.

In the gravity description,  the coupling to the reservoir is modeled by letting the matter CFT fields have transparent boundary conditions at the boundary of AdS.
After crossing the boundary of AdS,  they enter a non-gravitating flat space region.
This means that now the left and right movers are indeed independent and we will get both contributions from (\ref{spqleftright}).

Let the candidate QES be located at the point $P$ with $(t,z)$ coordinates equal to $(0,-a)$.
The generalized entropy function takes the form
\begin{align}
    \sgen(a) &= \frac{\phi_r}{a} + \left( 2\times \frac{c}{6} \log (b+a) - \frac{c}{6} \log \Omega(P) \right)\\
    &= \frac{\phi_r}{a} + \frac{c}{3} \log (a+b) - \frac{c}{6} \log a\, .
\end{align}
The extremization condition is
\begin{align}
    \frac{\phi_r}{a_\star} &= \frac{c}{6} \, \frac{a_\star-b}{a_\star+b}\, .
\end{align}
This is some quadratic equation that we can solve.
We record the answer for $b=0$,  when we get $a_\star = \phi_r \,  \frac{6}{c}$.

The ``surprise" is that,  for any value of $b$,  even if we don't include any portion of the wire,  the QES will be at some location away from the Poincar\'e horizon \cite{Almheiri:2019yqk}.
This means that the entanglement wedge of the complement system,  the reservoir, contains a disconnected piece far away from it,  near the horizon.
See the rightmost diagram in figure \ref{figzerotcoupled}.
Such disconnected pieces of an entanglement wedge have been dubbed ``islands" \cite{Almheiri:2019hni}. 
This is initially surprising because we might have (naively) argued that it is the SYK model which is dual to the AdS space,  so how can some part of AdS space map to the reservoir under the AdS/CFT map?
The point is simply that this argument is too naive,  and the full geometry is dual to the coupled SYK+reservoir system.
There is no paradox in having a disconnected entanglement wedge.

We have considered the case of fully reflecting and fully transparent boundary conditions for the matter fields at the AdS boundary.
If we take the matter to be free fermions, it is possible to perform the matter entropy computations for partially transmitting boundary conditions \cite{Kruthoff:2021vgv}.
As expected,  the QES interpolates between the fully reflecting and the fully transmitting cases.

\subsection{QES at nonzero temperature}
\begin{figure}
\centering
\begin{tikzpicture}

\filldraw[fill=gray!20] (0, 0) circle (0.3);
\draw (0.3,0) -- (2.3,0);
\fill (0.7, 0) circle (2pt);
\node[below right] at (0.7, 0) {$\mathsf{O}$};

\filldraw[fill=gray!20] (-1.5, 0) circle (0.3);
\draw (-1.8,0) -- (-3.8,0);
\fill (-2.2, 0) circle (2pt);
\node[below left] at (-2.2, 0) {$\mathsf{O}'$};

\draw[-{Latex[length=2mm]}] (-4,-2.5) -- (-4, 2.5) node[above]{$t$};

\end{tikzpicture}
\hspace{0.1in}
\begin{tikzpicture}

\draw (-2, -2.5) -- (-2, 2.5);
\draw (2, -2.5) -- (2, 2.5);

\draw (2, -2) -- (-2, 2);
\draw (2, 2) -- (-2, -2);

\draw (2, 2) -- (4, 0);
\draw (2, -2) -- (4, 0);

\draw (-2, -2) -- (-4, 0);
\draw (-2, 2) -- (-4, 0);

\node[below] at (0, -0.1) {$H$};
\node[left] at (2,2.1) {$A$};
\node[left] at (2,-2.1) {$B$};
\node[right] at (-2,2.1) {$A'$};
\node[right] at (-2,-2.1) {$B'$};
\node[right] at (4,0) {$C$};
\node[left] at (-4,0) {$C'$};

\fill (2.4, 0) circle (2pt);
\node[below] at (2.4,0) {$O$};

\fill (-2.4, 0) circle (2pt);
\node[below] at (-2.4,0) {$O'$};
\draw (2.4,0) -- (4,0);
\draw (-2.4,0) -- (-4,0);

\end{tikzpicture}

\caption{Left: We take two copies of the SYK model,  each coupled to a reservoir.  The right SYK+reservoir is in a thermal state.  The left SYK+reservoir acts as its canonical purification.
The ``radiation" region is defined to be the union of the reservoir sub-regions to the left of the point $\mathsf{O}'$ and to the right of the point $\mathsf{O}$.
We want to compute the entropy and entanglement wedge of this region,  this will be our model for the entropy of Hawking radiation in the eternal black hole setup.
We evolve the system with an operator that generates forward time evolution on both copies.
Right: The gravity dual of the setup.
The point $H$ is the bifurcation surface,  $AB$ and $A'B'$ are the boundaries of AdS,  and the regions $ABC$ and $A'B'C'$ are the non gravitating flat space regions.
}
\label{figtsolution}

\end{figure}
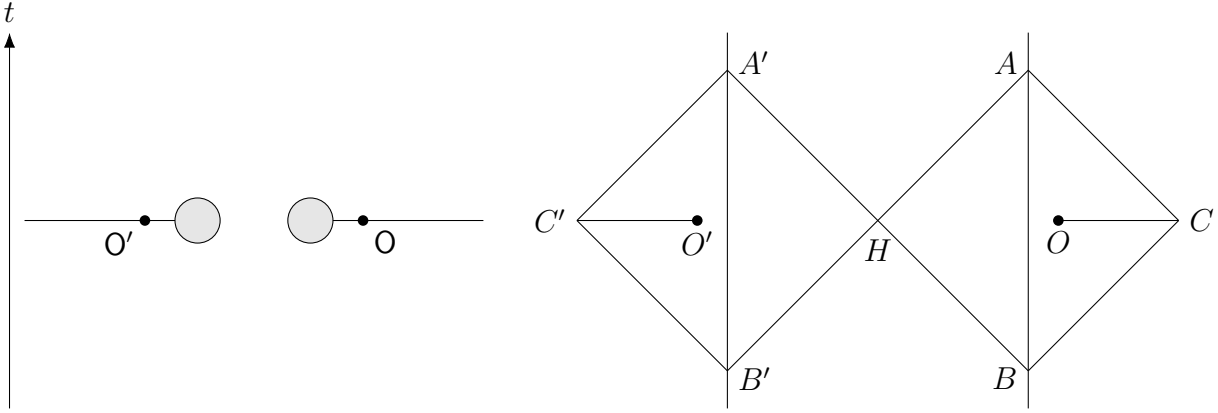

Let us now discuss the Euclidean solution of JT gravity that corresponds to a black hole with nonzero temperature.
The topology of a Euclidean 2d black hole is  that of a disk,  so the appropriate metric with $R=-2$ is given by
\begin{align}
    \d s^2 &= \d \rho^2 + (\sinh \rho)^2 \, \d \theta^2\, .
    \label{ds2disk}
\end{align}
Here $\rho=0$ is the horizon and $\rho = \infty$ is the boundary.
Let us see what the dilaton should be.
First simplify the equation (\ref{dilatoneom}) by taking the trace,  obtaining $\nabla^2 \phi = 2 \phi$,  and then plugging this back into (\ref{dilatoneom}),  yielding $ \nabla_a \nabla_b \phi = \phi \, g_{ab}$.
The left hand side is $\partial_a \partial_b \phi - \Gamma^c_{ab} \, \partial_c \phi$. 
When we look for a solution which is only a function of $\rho$, the $\theta\theta$ component gives us a first order equation.
It is $\frac{\phi'}{\phi} = \frac{\sinh \rho}{\cosh \rho}$,  which gives
\begin{align}
    \phi(\rho) = \phi_r \, \cosh \rho\,,
\end{align}
with the constant of proportionality being fixed by the boundary conditions.

The analytic continuation needed to get the Lorentzian solution is to make $\theta$ purely imaginary in (\ref{ds2disk}).
However,  let us first try to write the disk metric in a conformally flat form.
Define a coordinate $y$ using
\begin{align}
    \frac{\d \rho}{\sinh \rho} &=: \d y \,.
\end{align}
This implies that
\begin{align}
    \sinh \rho &= - \frac{1}{\sinh y}\, , \, \quad y \in (-\infty, 0)\\
    \cosh \rho &= - \frac{\cosh y}{\sinh y}\, .
\end{align}
So the metric (\ref{ds2disk}) becomes
\begin{align}
    \d s^2_\text{Euc} &= (\sinh y)^{-2} \left( \d y^2 + \d \theta^2 \right) \\
     \d s^2_\text{Lor} &= (\sinh y)^{-2} \left( \d y^2 - \d t^2 \right) \, . \label{ds2ty}
\end{align}
This is a good way to write the metric because we can obtain this via a simple reparametrization from the zero temperature solution.
Let us see this.
The zero temperature solution is
\begin{align}
    \d s^2 = - \frac{\d t^2 + \d z^2}{z^2} = - \frac{4}{(z^+ - z^-)^2} \, \d z^+ \, \d z^-\,,
    \label{ds2poincare}
\end{align}
where we have defined $z^{\pm} = t \pm z$.
If we transform to $(y^+,y^-)$ coordinates using $z^+ = \tanh \frac{y^+}{2}$ and $x^- =  \tanh \frac{y^-}{2}$,  we get
\begin{align}
    \d s^2 = - \left( \sinh \frac{y^+ - y^-}{2} \right)^{-2} \, \d y^+ \, \d y^-\,,
\end{align}
which is the same as (\ref{ds2ty}) with $y^{\pm} = t \pm y$.\footnote{This $t$ is different than the $t$ in (\ref{ds2poincare}).  This should not cause any confusion,  since from now on we will only use $t = \frac{y^+ - y^-}{2}$.}
Pictorially,  since $y<0$ in the AdS part,  we see that the $(y^+,y^-)$ coordinates cover the triangle $AHB$ in figure \ref{figtsolution},  with the horizon at $y = -\infty$ and the AdS boundary at $y=0$.
After we introduce the flat space bath,  it lies in the $y>0$ region and now the $(y^+,y^-)$ coordinates cover the diamond $AHBC$. 

The dilaton profile can be written using the $y$ variable as
\begin{align}
    \phi = \phi_r \cosh \rho = - \phi_r \coth y\, .
\end{align}
The point is that the $y$ coordinate is similar to the usual Poincare coordinate $z$ near the boundary. The dilaton behaves like $y^{-1}$ there.

In the previous discussion,  we set $\beta = 2\pi$,  but it will be useful to restore factors of $\beta$,  after which we get:
\begin{align}
    \d s^2 &= - \left(\frac{\beta}{2\pi}\, \sinh \frac{\pi}{\beta} (y^+ - y^-) \right)^{-2} \, \d y^+ \, \d y^- \,,\\
    \Omega &= \frac{\beta}{2\pi}\,\sinh \frac{2\pi (-y)}{\beta} \, , \quad y \in (-\infty, 0)\,,
    \label{omegay}\\
    \phi(y) &= \phi_r \, \frac{2\pi}{\beta} \,  \coth \frac{2\pi (-y)}{\beta} \,.
   \label{phiy}
\end{align}

Let us record the value of the black hole entropy.
It is given by
\begin{align}
    S_\text{BH}(\beta) = \phi_0 + \phi_r\, \frac{2\pi}{\beta} = \phi_0 + 2\pi \phi_r T\, .
    \label{sjtbh}
\end{align}
This linear in temperature behavior of the entropy is typical of near-extremal black holes.
The role of the area is played by value of the dilaton at a particular point.

The Lorentzian eternal two-sided black hole represents two copies of the dual quantum system in the thermofield double state.
As discussed in the introduction and in the beginning of this section,  we would like to couple each of these two systems to a reservoir and consider the setup shown on the left in figure \ref{figtsolution}.
We would like to compute the entropy of the ``radiation region",  which is the union of the region to the left of the point $\mathsf{O}'$ and the region to the right of the point $\mathsf{O}$ in the left panel of figure \ref{figtsolution}.
Recall that we are evolving the system with a generator that evolves both systems forwards in time.
The thermofield double state is a quenched state with respect to this Hamiltonian,  so we expect the entropy to initially increase linearly in time.
In the next subsection,  we calculate this linear increase.

\subsubsection{Early time linear growth}
The early time linear growth is a purely CFT computation.
We compute the entropy of the complement of the radiation system,  which is the interval $OO'$.

We imagine that the points $O$ and $O'$ in the right panel of figure \ref{figtsolution} have been evolved forward by time $t$.
The state of the matter CFT is thermal when restricted to the diamond $AHBC$.
Thus,  we have to import some results from section \ref{sec:cftentropy}. 
In particular,  we have to introduce $(x^+,x^-)$ coordinates via Eq.(\ref{xyrelation}).
 \begin{align}
    \begin{tabular}{c|c|c}
    & $O$ & $O'$ \\
    \hline
      $(t,y)$     & $(t,b)$ &  \\
      $(y^+,y^-)$  & $ (t+b,t-b) $ &  \\
       $(x^+,x^-)$  & $ (e^{t+b}, - e^{-t+b}) $ & $(- e^{-t+b}, e^{t+b})$ \\
    \end{tabular}
\end{align}
We do not write the $(t,y)$ or the $(y^+,y^-)$ coordinates of the point $O'$ since this point lies outside the region that is covered by these coordinates.
The $(x^+,x^-)$ coordinates of $O'$ are related to those of $O$ by reflection in the vertical line $x^+ = x^-$,  which simply swaps the two coordinates.

Thus,  the coordinate separation between $O$ and $O'$ is $\Delta x^+ = e^b \cdot 2 \cosh t$.
The separation along $x^-$ is the same.
For both points, $\Omega = (-x^+ x^-)^\frac{1}{2} = e^b$.
Thus,  using the formula (\ref{spqomega}),  we get
\begin{align}
    S_\text{rad}(t) &= 2\cdot \frac{c}{6} \log (e^b \cdot 2 \cosh t) - \frac{c}{6} \log e^b - \frac{c}{6} \log e^b \\
    &= \frac{c}{3} \log (2 \cosh t) \approx \frac{c}{3}\,  \frac{2\pi t}{\beta}\, ,
    \label{sradt}
\end{align}
where we restored $\beta$ in the last expression after taking the limit $t \gg \beta$.
Thus,  we get a linear-in-time increase in the entropy.

Now,  if this linear increase were to continue forever,  we would run into an entropy paradox,  which is very similar to Page's: the combined system is in a pure state and the complement system (which for us is the union of the two SYK models) has bounded entropy.
We will now see how disconnected entanglement wedges save us from this entropy paradox.

\subsubsection{Late time saturation and islands}
\begin{figure}
\centering

\begin{tikzpicture}

\draw (-2, -2.5) -- (-2, 2.5);
\draw (2, -2.5) -- (2, 2.5);

\draw (2, -2) -- (-2, 2);
\draw (2, 2) -- (-2, -2);

\draw (2, 2) -- (4, 0);
\draw (2, -2) -- (4, 0);

\draw (-2, -2) -- (-4, 0);
\draw (-2, 2) -- (-4, 0);

\node[below] at (0, -0.1) {$H$};
\node[left] at (2,2.1) {$A$};
\node[left] at (2,-2.1) {$B$};
\node[right] at (-2,2.1) {$A'$};
\node[right] at (-2,-2.1) {$B'$};
\node[right] at (4,0) {$C$};
\node[left] at (-4,0) {$C'$};


\draw[red] (-2.3,1) -- (-4,0);
\draw[red] (2.3,1) -- (4,0);
\draw[red] (-1,0.5) -- (1,0.5);

\fill (-2.3,1) circle (2pt);
\node[below] at (-2.3,1) {$O'$};
\fill (2.3, 1) circle (2pt);
\node[below] at (2.3,1) {$O$};

\fill (-1,0.5) circle (2pt);
\node[below] at (-1,0.5) {$P'_\star$};
\fill (1, 0.5) circle (2pt);
\node[below] at (1,0.5) {$P_\star$};


\end{tikzpicture}

\caption{The entanglement wedge of the radiation region contains a disconnected region $P_\star P'_\star$ at late times,  which is called the island.
The entropy computed using this QES does not grow with time.  In fact,  it almost equals the sum of the thermodynamic entropy of the two black hole,  a constant.
}
\label{figtisland}

\end{figure}
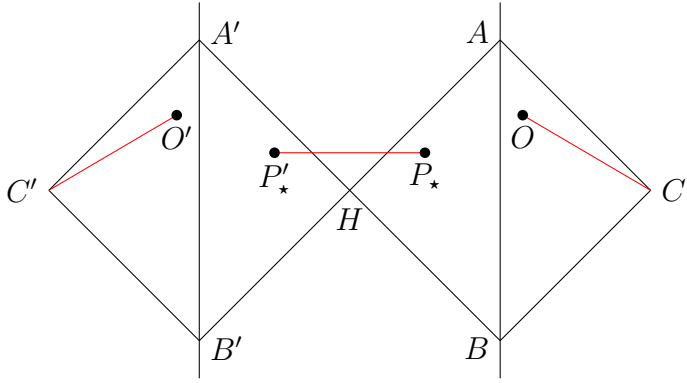
The claim is that the entanglement wedge of the radiation region contains a disconnected ``island" region that spans the interior of the black hole.
So,  as a candidate QES configuration,  we should compute the generalized entropy function of a region consisting of two intervals $O'P'$ and $OP$.
Since two-interval entropies depend on the details of the CFT,  we use a further approximation that,  at late times,  the answer will just be a sum of the entropies.\footnote{Technically speaking,  the four-point function of twist operators in this limit is dominated by an OPE expansion,  and only the identity operator contributes in the intermediate channel.}

So we consider computing the entropy of just the right SYK system plus a piece of the reservoir of length $b$.
We construct the generalized entropy functional for the region $OP$,  where the point $P$ has coordinates $(t,y) = (0,-a)$,  see figure \ref{figtisland}.
It takes the form
\begin{align}
    \sgen(a) =  \phi_r \, \frac{2\pi}{\beta} \, \coth \frac{2\pi a}{\beta} + 
    \frac{c}{3} \log \left( \frac{\beta}{\pi} \sinh \frac{\pi}{\beta} (b+a) \right)
    - \frac{c}{6} \log \left(\frac{\beta}{2\pi}\sinh \frac{2\pi a}{\beta}\right)
    \label{sgenpeninsula}\, .
\end{align}
The first term in this formula is the dilaton (\ref{phiy}),  the second term is the thermal entropy of the interval $OP$,  given by (\ref{sthermal}),  and the final term is the correction due to the Weyl factor (\ref{omegay}) at the point $P$.
Setting the derivative of equal to zero,  after some simplification we find the QES condition
\begin{align}
    \phi_r \, \frac{2\pi}{\beta} \, \frac{1}{\sinh \frac{2\pi a_\star}{\beta}}
    &= \frac{c}{6} \, 
    \frac{\sinh \frac{\pi}{\beta}(a_\star-b)}{\sinh \frac{\pi}{\beta}(a_\star+b)}\, .
    \label{qescondition}
\end{align}

Let us solve this equation in the slow emission/absorption limit,  which is
\begin{align}
    \frac{\phi_r}{\beta} \gg c\,.
\end{align}
This condition says that the entropy of the black hole above extremality (\ref{sjtbh}) is large compared to the entropy that it will lose or gain per unit thermal time due to emission and absorption of radiation quanta.
In this limit, we can approximate all $\sinh$'s in (\ref{qescondition}) with exponentials which simplifies the equation to
\begin{align}
    \log \left( \phi_r \frac{2\pi}{\beta} \frac{6}{c}  \right) &+ \log 2 - \frac{2\pi a_\star}{\beta} \approx - \frac{2\pi b}{\beta}  \,,\quad \text{ or }\\
    a_\star &\approx \frac{\beta}{2\pi} \log \left( \phi_r \frac{2\pi}{\beta} \frac{12}{c}  \right) + b\, .
\end{align}
Here $b$ itself could be small or large,  we could just set it zero.
Since $\frac{\phi_r}{\beta c} \gg 1$,  we have $a_\star \gg 1$,  which just says that the QES is close to the horizon.
The actual value of the entropy at this extremum is approximately given by the first term in (\ref{sgenpeninsula}) and with the $\coth$ replaced by one.\footnote{Dividing by $c$,  the first term is order $\frac{\phi_r}{\beta c}$ and the others are of order $\log \frac{\phi_r}{\beta c}$.}
This is just the black hole entropy (\ref{sjtbh}).
(We dropped by constant $\phi_0$ from our expression for $\sgen(a)$ since it does not contribute to the extremization condition,  but it is really present and we should add it back.)
Thus,  we conclude that
\begin{align}
S_\text{rad}(t) \approx 2 S_\text{BH}(\beta)\quad \text{for }  \frac{c t}{\beta} \gg S_\text{BH} \, .
\end{align}
The factor of two on the RHS comes from having two intervals $OP_\star$ and $O'P'_\star$.

To summarize,  the entanglement entropy of the radiation region initially grows linearly in time (\ref{sradt}).
At a time of order $\beta \frac{S_\text{BH}}{c}$,  this linear growth is halted by the appearance of a disconnected portion of the entanglement wedge of the radiation.
This disconnected portion lies in the gravitating region and spans the interior of the two-sided black hole,  see figure \ref{figtisland}.
The entropy computed using the QES formula at late times agrees with the predictions from general principles of unitary time evolution.
This early time increase,  followed by the late time saturation is the analog of the Page curve for the eternal black hole.

\section{Concluding remarks}
The fact that the entanglement wedge of the radiation region, which lies in a region of weak gravity,  contains a part of the black hole interior often causes major discomfort.
One may start worrying about ``nonlocal effects" and some ``physical transport of information" happening from the island region to the radiation region.
Nothing of the sort is being claimed.

Let us recap the logic.
We first decide which quantity $Q$ on the boundary we want to compute.
For the Page curve computations,  it is the von Neumann entropy,  or perhaps the Renyi entropies.
Then one sets up the boundary path integral that is relevant for computing the quantity $Q$.
Then one uses AdS/CFT to compute this path integral via gravitational saddle-points.
There might be multiple competing saddle-points,  and a saddle-point that is initially suppressed might become dominant when some parameter is varied.
This is the full logic of the calculation,  and it fits within very standard Gibbons-Hawking and AdS/CFT computations.
If one is worried about ``how the information is getting out",   one should first formulate a precise version of this question on the boundary and identify a quantity that one wants to compute. 
Then one should set up the path integral on the boundary and compute the quantity using gravitational saddle points.

In this context,  it seems like an interesting question to the author to ask what computation on the boundary captures the Hawking particle creation process near the black hole horizon.

We would like to stress again that the QES formula or the ``island prescription" is not fundamental. 
It is the above path integral setup that is fundamental.
In the table below,  we record three entropy ``prescriptions" in the left column and their corresponding derivations using the boundary path integral logic.
The entropy ``prescriptions" work in the original geometry when the problem under study is simple enough.
The right column should be treated as more fundamental.
\begin{center}
\begin{tabular}{|c|c|c|}
\hline
1 & Bekenstein-Hawking area formula & Gibbons-Hawking path integral\\
\hline
2 & Ryu-Takayanagi formula & Lewkowycz-Maldacena\\
\hline
3 & QES formula with islands & Boundary path integral with replica wormholes\\
\hline
\end{tabular}
\end{center}

Let $\rho$ be the density matrix of Hawking radiation.
Note that $-\Tr(\rho \log \rho)$ is just one real number made out of an exponentially large number of matrix entries of $\rho$.
The surprise of the papers \cite{Almheiri:2019psf,  Penington:2019npb,  Penington:2019kki, Almheiri:2019qdq} is that this quantity happened to be computable using semi-classical gravity,  albeit using some nontrivial saddle points.
It is also a surprise,  hailing back to the 1970s,  that semi-classical gravity knows about the entropy of the black hole.
It is a more recent surprise that semi-classical gravity knows something about the chaotic nature of the eigenvalues of the boundary Hamiltonian \cite{Saad:2018bqo}.

However,  it is also clear that semi-classical gravity cannot know all the details of the microscopic boundary answers.
For example,  it does not seem to know about the large erratic fluctuations that exist on top of an averaged ``ramp" feature in the spectral form factor or real-time thermal two-point functions \cite{Cotler:2016fpe}.
It also does not seem to know about the individual exponentially large number of matrix elements of the full density matrix of Hawking radiation.
That's a good thing. 
We know,  after all,  that Einstein gravity suffers from UV problems and, so semi-classical GR coupled to matter is definitely not the ultimate theory of quantum gravity.
However,   it is still an interesting question to ask:  What other subtle quantities of the boundary quantum theory can be computed using semi-classical gravity,  improved by wormholes?

\paragraph{Acknowledgments.} 
I would like to thank the organizers of the Asia Pacific Center for Theoretical Physics online winter school held in January 2021 and the organizers of the Quantum Information, Quantum Field Theory and Gravity program held at ICTS,  Bengaluru in August 2024 for inviting me to lecture about these topics.

\bibliographystyle{apsrev4-1long}
\bibliography{bh_info_chapter}
\end{document}